**Enhancing Light Emission in Interface Engineered Spin-OLEDs Through Spin-Polarized Injection at High Voltages**

*J.P. Prieto-Ruiz,[a#] S.G. Miralles,[a#] H. Prima-García,*[a] A. López-Muñoz,[a] A. Riminucci,[b] P. Graziosi,[b] M. Aeschlimann,[c] M. Cinchetti,[d] V. A. Dediu,*[b] E. Coronado*[a]*

[a] Instituto de Ciencia Molecular (ICMol), Universidad de Valencia. Catedrático José Beltrán 2, 46890 Paterna, Spain. Fax: +34 96 354 3273. Telf: +34 96 354 4419. E-mail: eugenio.coronado@uv.es, helena.prima@uv.es

[b] Instituto per lo Studio dei Materiali Nanostrutturati ISMN - CNR, Via Gobetti, 101, Bologna, 40129, Italy. E-mail: V.Dediu@bo.ismn.cnr.it

[c] Department of Physics and Research Center OPTIMAS, University of Kaiserslautern, Erwin-Schroedinger-Strasse 46, 67663 Kaiserslautern, Germany.

[d] Experimentelle Physik VI, Technische Universität Dortmund, 44221 Dortmund, Germany.

#These authors have equally contributed

The quest for a spin-polarized organic light emitting diode (spin-OLED) is a common goal in the emerging fields of molecular electronics and spintronics. In this device two ferromagnetic electrodes are used to enhance the electroluminescence intensity of the OLED through a magnetic control of the spin polarization of the injected carriers. The major difficulty is that the driving voltage of an OLED device exceeds of a few volts, while spin injection in organic materials is only efficient at low voltages. We report here the fabrication of a spin-OLED that uses a conjugated polymer as bipolar spin collector layer and ferromagnetic electrodes. Through a careful engineering of the organic/inorganic interfaces we have succeeded in obtaining a light-emitting device showing spin-valve effects at high voltages (up to 14 V). This has allowed us to detect a magneto-electroluminescence enhancement on the order of a 2.4 % at 9 V for the antiparallel configuration of the magnetic electrodes. This observation provides evidence for the long-standing fundamental issue of injecting spins from magnetic electrodes into the frontier levels of a molecular semiconductor. Our finding opens the way for the design of multifunctional devices coupling the light and the spin degrees of freedom.

Molecular spintronics has successfully crossed its first 15 years of life becoming an important player in the race towards the fabrication of next generation spintronic devices.[1–5] In this context, the development of multifunctional spintronic devices, taking advantage of the properties exhibited by molecular semiconductors, represents one of the key challenges in this area.[1,6–10] An archetypical example of this type is provided by so-called spin-OLEDs, in which spin-polarized injection is combined with light emission.[7,8,11,12] In these multifunctional devices the spin polarization (SP) is expected to establish a quantum control over the statistics of singlet and triplet excitons in the emitting material, changing the singlet-to-triplet population and enabling a magneto-electro-optical multifunctionality.[6,13] This may represent a new route for magnetically tuning the light emission and, more importantly, it may embody a powerful tool for the demonstration of induced SP in the electronic states (HOMO and LUMO) of an organic material. In fact, the possibility to establish non-equilibrium SP inside a molecular material is still an open question.[2,7–9,14,15] We tackle this fundamental problem in a specially designed OLED with spin-polarized injectors.

The first devices of this kind, fabricated without deep interface engineering, required excessively high voltages for the generation of a detectable light intensity.[16] Further interface improvements solved this problem, opening the possibility for a real analysis of the SP effects. In this context two papers stand out. Salis *et al.*[17] reported a Ni/SAD/Alq$_3$/Py OLED and did not detect any effect coming from the SP of the injected carriers. Instead, this study revealed an important artefact: the stray fields of these electrodes cause the so-called organic magnetoresistance (OMAR) effect in the molecular layer.[18] Since the stray field depends on the mutual orientation of the electrodes, this situation mimics the one usually associated with spintronic effects. In a more recent paper, Nguyen *et al.* fabricated a magnetic OLED based on LSMO/DOO-PPV/LiF/Co with the aim of achieving a magnetic control in the electroluminescence (EL) of the device by means of the SP injecting electrodes.[12] Aware of the artefacts described above, they demonstrated the detection of magnetoresistance (MR) and especially the detection of a small but clearly measureable decrease in the EL for the antiparallel orientation of the spin polarised electrodes compared to the parallel one. The employed detection scheme, consisting in EL measured at constant voltage, left unfortunately ambiguous the achievement of the spintronic control of these OLEDs. Indeed, the detected reduction of light emission occurred simultaneously with the current reduction through the device caused by positive magnetoresistive effect. Thus, the possibility that the effect could be due to the trivial reduction of exciton pumping rate cannot be ruled out.

We present below a highly efficient OLED with spin polarized electrodes, in which the manipulation of the spin dependent exciton statistics is unambiguously controlled by the mutual magnetic orientation of the electron and hole spin polarized injectors. We fabricate a spin-polarized OLED device based on a hybrid organic-inorganic LED (HyLED).[19] We have chosen such a structure since its driving voltage, $V_{on}$, has shown to be comparatively lower than that provided by a conventional OLED. The structure of the device is sketched in Figure 1a and the corresponding HRTEM image of the cross section of the device showing the quality of the interfaces is depicted in Figure 1b. As FM electrodes, we have used LSMO (cathode) and Co (anode) since they have work functions close to those of the non-magnetic electrodes commonly used in HyLEDs (ITO and Au, respectively). As organic semiconductor, the light emitting conjugated polymer poly(9,9-dioctylfluorene-co-benzothidiazole) (F8BT) has been used since it can provide intense green EL (Figure S1-S2); in addition, its bipolar character can facilitate the establishment of a bipolar charge transport regime in the device.[19,20]

To optimize the charge injection from the electrodes to the F8BT, interfacial layers have been inserted to tailor the energy-level alignment between the molecule's frontier orbitals and the electrode's work functions.[21] We have used ultraviolet photoemission spectroscopy (UPS) to guide the choice of appropriate interfacial layers (see Methods and Figure S3). Figure 1c shows the energy level alignment of the LSMO/F8BT/Co device without engineered interfaces, while Figure 1d shows the energy level alignment of the interface-engineered device. In this device we have inserted a thin layer (3 nm) of molybdenum oxide ($MoO_x$) at the Co/F8BT interface to move the HOMO level of the F8BT layer close to the Fermi level, and thus facilitate holes injection. $MoO_x$ has been extensively used as hole injection layer (HIL) in hybrid OLED devices.[19] In a recent work we have shown by spin-polarised UPS that $MoO_x$ increases the work function of the cobalt electrode from -5.0 to -5.8 eV, thus reducing the hole injection barrier from the electrode to the HOMO of the F8BT.[13] More interestingly, up to a nominal $MoO_x$ thickness of 3 nm, the spin polarization close to the Fermi level is virtually unaffected by the presence of this interfacial barrier.

On the other hand, to move the LUMO level of the F8BT close to the Fermi level of the LSMO electrode, and thus to reduce the electron injection barrier, we have used the commercial polymer polyethylenimine ethoxylated (PEIE; Figure S1a) as electron injection layer (EIL). PEIE contains simple aliphatic amine groups able to create a strong interface of molecular dipoles. This surface modifier has shown to substantially reduce the work function of the metallic electrodes in organic optoelectronic devices, thus facilitating the injection of electrons to the

LUMO of the organic semiconductor.[22] In the present case, UPS measurements have shown that the deposition of an ultrathin layer (c.a. 1 nm) of PEIE on LSMO through spin coating reduces the work function from -4.7 to -3.7 eV (Figure 1d and Figure S3). This value matches well with that of the LUMO of the F8BT (-3.5 eV). As optoelectronic device, this hybrid LED structure exhibits at low temperatures the EL spectrum characteristic of F8BT, which is maintained in all the range of temperatures investigated (Figure S2). The effect of the EIL on the EL of the device can be clearly seen in Figure 2: at low temperatures (20 K) the presence of PEIE significantly decreases the $V_{on}$ value from 10 V to 7 V, while the luminance increases by more than one order of magnitude. As expected, due to the semiconducting character of F8BT, this driving voltage is higher than that measured at room temperature ($V_{on}$= 3.7 V).

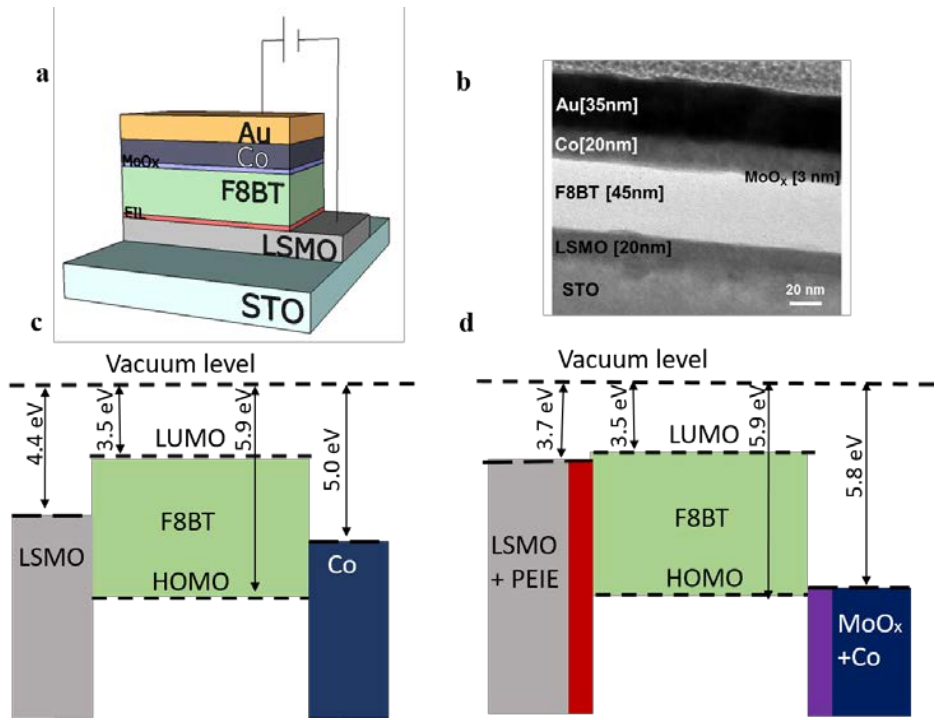

**Figure 1. Sketch of the structure of the spin-OLED device showing the band energy levels**. **a**. Schematic structure of the layers composing the device. **b**. HRTEM image of the cross section

of the device showing the high quality of the interfaces. **c.** Energy band alignment of the device without using interfacial layers between the organic semiconductor (F8BT) and the electrodes. **d**. Energy band alignment in the interface-engineered device.

Note that this hybrid structure is very robust from the optoelectronic point of view, as it maintains the EL features when PEIE is changed by another EIL. To support this point we have used as EIL the Ru(II) complex bis(4,4'-tridecyl-2,2'-bipyridine)-(4,4'-dicarboxy-2,2'-bipyridine)ruthenium(II)-bis(chloride) (in short N965; Figure S1b). This amphiphilic molecule has already been used for this purpose in hybrid LED structures.[13] In the present example it has been organized as a monolayer on LSMO using the Langmuir-Blodgett technique. UPS measurements show that the presence of this dipolar monolayer leads to a reduction in the work function of the LSMO similar to that obtained for PEIE (from -4.7 to -3.8 eV, Figure S3). As a result, this EIL also leads to a similar improvement in both $V_{on}$ value and luminance intensity (Figure 2, Figure S4, Figure S13).

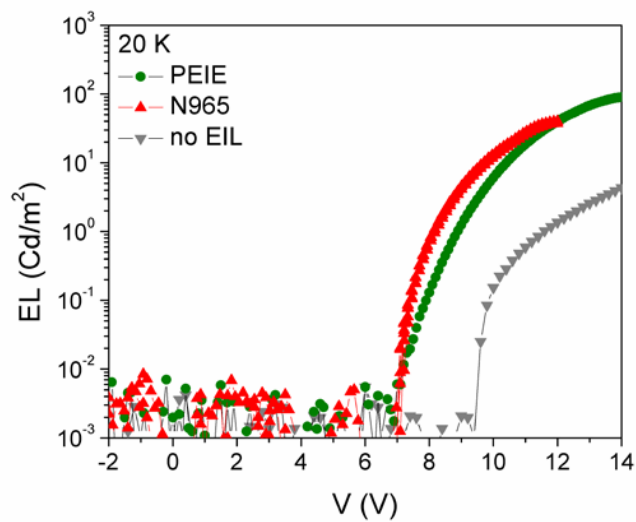

**Figure 2. Electroluminescence of the spin-OLED device upon insertion of an EIL.** Data recorded at 20 K varying the bias voltage from -2 to 14 V. For both, the device fabricated with PEIE (green circles) and that fabricated with N965 molecules (red up-pointing triangles), $V_{on}$= 7 V, while for the device without EIL (grey down-pointing triangles) this value increases to $V_{on}$= 9.5 V.

As far as the spintronic properties of the multifunctional device are concerned, we observe that the field dependence of the resistance exhibits the typical features of a spin valve, indicating that the current passing through the device is able to preserve its spin-polarization. Thus, two different resistance states are observed depending on the relative magnetic alignment of the two magnetic electrodes. Figure 3a shows the MR of the device measured at 20 K under a voltage of 9 V. The MR of the device at low voltages is showed in Figure S11. This MR response is hysteretic and follows the coercive fields of the electrodes (see Figure S5). It is remarkable that this spin valve response is maintained up to voltages of 14 V, which are well above the driving voltage of the OLED (Figure 3b). The SP current is preserved up to 180-200 K, with a thermal evolution of the MR resembling the one of the bulk LSMO magnetization (Figure 3c and Figure S6c).

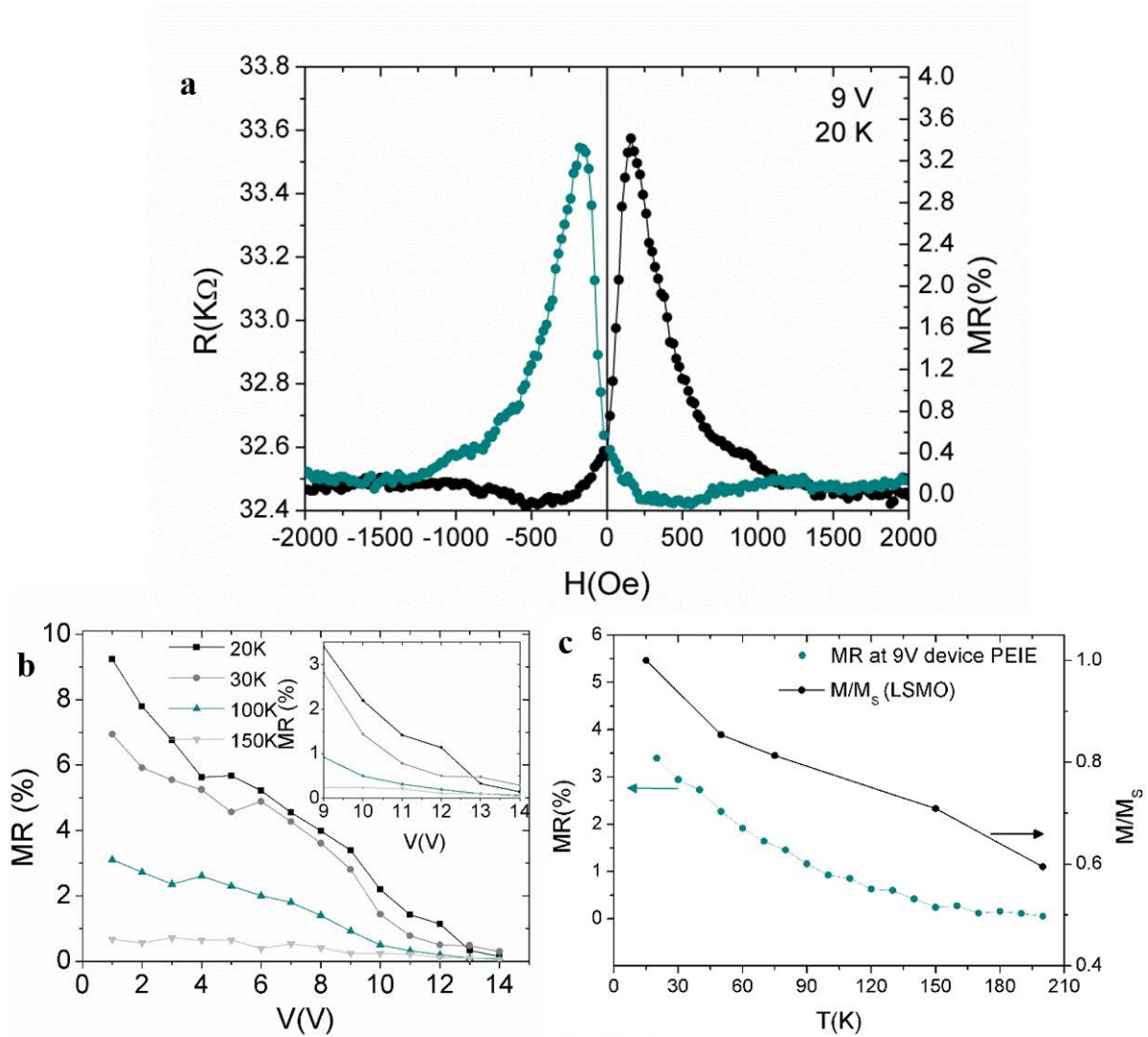

**Figure 3. Magnetoresistance characterization of the spin-OLED device. a**. MR at 9 V and 20 K of the device containing PEIE as EIL. Purple and grey arrows point the FM electrodes magnetization direction. **b**. MR as a function of voltage for different temperatures. **c**. MR at 9 V and LSMO magnetization decay as a function of temperature. The absence of a flat antiparallel region is quite typical behaviour in spintronic devices, even for inorganic materials.[23,24]

The stabilization of a SP current in the same range of voltage and temperatures, where light emission is observed, is the necessary condition to detect a MEL response in the hybrid device. The MEL response is defined by (($EL_{ap}$-$EL_p$)/$EL_p$), being $EL_{ap}$ and $EL_p$ the experimental electroluminescence values obtained in the antiparallel and parallel states of the electrode's magnetization, respectively.[12] In the present case, MEL is indeed observed showing a maximum value of 2.4 % in the AP state at 9 V and 20 K (Figure 4a). The absolute value of the MEL effect for different voltages (Figure S8) follows the variation of the MR with voltage (Figure 3b). It is observed for voltages above 8 V. Furthermore, it decreases when the temperature is increased being clearly observed up to 100 K.

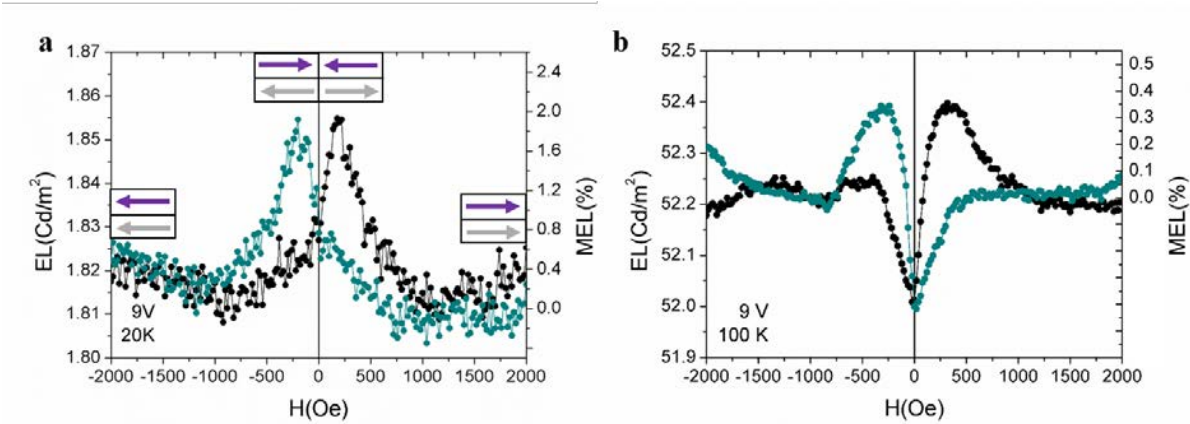

**Figure 4. Magnetoelectroluminescence (MEL) characterization of the spin-OLED device. a**. MEL response at 9 V and 20 K exhibiting a maximum enhancement of light emission of 2.4 % in the AP state configuration of the electrode's magnetization. **b**. MEL response measured at 100 K and 9 V showing the appearance of an important OMEL contribution due to the organic F8BT polymer.

This MEL effect gives rise to an enhancement of the EL in the AP state, in agreement with the presence of spin-polarised charge carriers in the F8BT. At temperatures close to 100 K the magnetoelectroluminescence is masked by the intrinsic organic MEL effect (OMEL, Figure 4b). This last effect is related to the external magnetic field and not to mutual magnetisation of the electrodes being opposite to the previous one and leading to a minimum value of the EL at zero field. It originates from the influence of variations in the local magnetic field over the recombination and transport processes of the injected carriers in the organic semiconductor.[25,26] The hyperfine nuclear field in the organic semiconductor is the major responsible of this effect. As the temperature is progressively increased, the OMEL effect is enhanced at expenses of the MEL signal related to the spin valve performance of the device, which sharply diminishes upon heating, being therefore masked by the OMEL signal above ca. 100 K. This evolution results from the weak temperature dependence of the OMEL effect, directly observed in a blank sample formed by non-magnetic electrodes (see Fig. S12), and also pointed out by other authors in other semiconductor thin film devices.[27,28]

The fact that the MEL response follows the coercive fields of the electrodes (Figure S5) is a first indication that a SP is the responsible of this behaviour. However, a systematic study has been performed in order to discard that this light modulation comes from artefacts induced by the electrodes in the structure, which could mimic spin dependent processes of the injected carriers. The first artefact is the coupling of the current with light emission. According to this, an enhancement of the light emission should be observed when the current through the device is increased. In our case we observe the opposite trend (i.e., an enhancement of the light emission when the current is reduced by the spin valve effect in the AP state, Figure S9). This result discards the artefact and supports the SP effect as responsible for the light modulation[7,16,29] in clear contrast with previous reports.[12]

A second artefact may be induced by the stray fields of the ferromagnetic electrodes.[17] This can induce a hysteretic response of the EL under sweeping of an external magnetic field, which would not be related to the spin valve effect in the device. In order to discard this situation three reference devices were fabricated (without ferromagnetic electrodes, with Co as single FM electrode and with LSMO as single FM electrode) and studied in the same conditions as the spin-OLED (Figure S10). In the three cases no hysteretic features were observed and the EL response had an opposite sign with respect to the MEL effect observed in the spin-OLED.

A third artefact may be related with the effects of the ferromagnetic electrodes on the current and light emission responses under the application of external magnetic fields. We have discarded possible artefacts in the measured MR and MEL signals caused by high resistances of the ferromagnetic electrodes because the resistance of the Co can be neglected and LSMO cannot induce any artefact since the device is 10 times more resistive than this electrode and furthermore its MR shows the same sign as the device.

The above discussion clearly demonstrates the achievement of spin-polarised injection in the electronic levels of the organic semiconductor. This demonstration has been enabled by the ability to maintain this SP at high bias voltages. This feature is quite unusual for molecular spin valves,[7,15,30,31] which, in general, show a complete degradation of their spin-polarization signal for high voltages. As mentioned before, one of the reasons is that for most molecular spin valves the spin injection does not take place through the frontier orbitals of the organic semiconductor, but through its traps or impurities.[14] In our case, we have shown that, through a careful interface engineering, the energy-level alignment between the molecule's frontier orbitals and the electrode's work functions can be optimized in such a way that polarised charges are injected in an efficient way at the HOMO and LUMO levels of the F8BT, thus maintaining the spin-polarization at high voltages. This result is also favoured by the fact that in our device these charges recombine to emit light for $V>V_{on}$. Thus, at these high voltages this recombination of charges leads to a significant reduction of charge density in the F8BT. Such a mechanism may contribute to reduce spin-spin scattering processes, leading to a reduction of this spin relaxation rate.[32] The understanding of the particular transport mechanism in the framework of HOMO/LUMO channels, on the other hand, is a difficult and laborious task and requires a dedicated research going well beyond the goals of this paper. For example, details of transport in some organic materials are still debating. Noteworthy, to the best of our knowledge this is the first example of devices showing both SP related MEL and magnetic field related OMEL effects (see above), reversibly enhanced or suppressed by voltage variations. This coexistence represents and additional evidence for carrier transport along the HOMO/LUMO levels.[33]

It is worth to note that this hybrid structure is not only robust from the point of view of its optoelectronic properties, but also from the point of view of the spintronic properties. In fact, the spin polarized properties are maintained when the interfacial EIL layer is changed from PEIE to N965. (Figure S4-S5-S6-S7-S10). In general, the performance of the N965-based device shows lower values than the PEIE-based device in both the MR (6 % compared to 9.5 % at 20 K and 2 V) and the maximum enhancement of the MEL (1 % compared to 2.4%). A possible reason to explain this difference may be related with the stronger spin-orbit coupling of the N965 molecule, as it contains Ru.

In summary, we have prepared a robust spin-OLED that shows an increase in the magneto-electroluminescence (≈ 2.4 %) for the antiparallel configuration of the magnetic electrodes. This result is especially relevant since for this configuration the total current flowing through the OLED is reduced with respect to that for the parallel configuration due to magnetoresistance effects. In OLEDs the singlet/triplet ratio is determined by quantum ich statistics to be 1:3. In an ideal spin polarized OLED with a 100 % spin polarization. This ratio increases up to 1:1 for opposite spin polarization of injected electrons and holes, thus leading to a significant increase of EL efficiency. In contrast, parallel spin states for electrons and holes result in the exclusive formation of triplets, which are not emitting and therefore the EL vanishes. In our device, these two limit values are not reached since only a small fraction of the injected charges are spin polarized. In order words, the polarized spin current represents a small effect of the total current and consequently only a small fraction of the measured EL is caused by spin polarization. This makes difficult to use MEL results to directly estimate the amount of spin polarization. In any case, the enhanced MEL effect observed in our device provides an unequivocal evidence for the long-standing fundamental issue of inducing spin polarization in the bulk of molecular solids. Noteworthy, this proof, together with the absence of the Hanle effect in molecular spintronic devices —well-established previously and also confirmed for these devices— evokes a conceptually new physics for the description of the spin-environment interactions in molecular solids. This novel strategy, which employs a hybrid organic-inorganic LED structure and commercial materials to design a robust spin-OLED, opens the way for preparing new types of molecular-based multifunctional spintronic devices showing a synergy between spin and a second functionality such as light emission, electrical memory or photovoltaics.

**Experimental Section**

*Fabrication of spin-OLEDs and materials*. The device under study was fabricated by combining solution processing methods and evaporation techniques. A LSMO electrode with a thickness of 20 nm was grown by Channel-Spark ablation (CSA)[34] on (100) oriented transparent SrTiO3 substrates, and employed as the cathode and spin injector electrode of the structure. For the device with PEIE as EIL, Polyethylenimine, 80% ethoxylated (PEIE) (Mw = 70,000 g/mol), was dissolved in H2O with a concentration of 35-40 wt.% as received from Aldrich. Then it was further diluted with 2-methoxyethanol to a weight concentration of 0.025%. The solution was spin coated on top of the substrates at a speed of 5000 rpm for 1 min and an acceleration of 1000 rpm/s. Spin-coated PEIE films were annealed at 100 ºC for 10 min on a hotplate in ambient air. The thickness of these PEIE layers was determined to be ca. 1 nm by combining AFM scratching technique and absorption measurements. For the device with N965 as EIL, a molecular ionic junction constituted by a monolayer of a charged ruthenium complex[35] was deposited by Langmuir-Blodgett technique on the LSMO cathode. The synthesis of this ionic ruthenium complex, with formula bis(4,4'-tridecyl-2,2'-bipyridine)-(4,4'-dicarboxy-2,2'-bipyridine)ruthenium(II)-bis(chloride) (N965), is described elsewhere.

After the deposition of the EIL, a 45 nm thick film of the light emitting conjugated polymer poly(9,9-dioctylfluorene-co-benzothidiazole) named as F8BT, received from Aldrich, was deposited by spin-coating from a chlorobenzene solution. After the polymer deposition, the sample was transferred to a thermal evaporator where 3 nm $MoO_X$ layer were evaporated on top of the whole device. This thickness is a good compromise for the correct optoelectronic performance of the device as well as from the spintronic point of view.[13] A new shadow mask in a cross-bar configuration respect to the LSMO was used for the deposition of 18 nm of Co by thermal evaporation. Without breaking the vacuum, a gold capping layer of 35 nm was deposited. The active area amounts to 500 x 500 µm.

*Magnetoresistance and magnetoelectroluminescence measurements*. The MR ratio in the spin-OLEDs has been calculated taking the parallel configuration of electrode magnetizations as the reference, using the expression $\Delta R/R_p = ((R_{ap}-R_p)/R_p) \cdot 100\%$ where $R_p$, and $R_{ap}$, the resistance in the parallel and antiparallel states, respectively. The LSMO contribution in the MR/MEL responses has been systematically subtracted from each experimental curve. The magnetoelectroluminescence intensity was detected by employing a Si-photodiode coupled to a Keithley 6485 picoamperometer. The photocurrent was calibrated using a Minolta LS100 luminance meter. An Avantes luminance spectrometer was used to measure the EL spectrum of the device.

*Work function measurements.* To detect the work function at the engineered interfaces (LSMO/PEIE, LSMO/N965) we have used UPS in a setup similar to the one described in the work23. Shortly, the excitation source is a VUV5000 monochromatized vacuum ultraviolet lamp (Scienta), operated at the He II line (hv = 40.8 eV). The light incident angle is 45◦. The photoelectrons are collected at normal emission using a hemispherical energy analyser for parallel energy and momentum detection (SPECS Phoibos 150). All presented measurements were performed at room temperature.

**Acknowledgements**
Financial support from the EU (COST Action MOLSPIN CA15128, FET-OPEN 2D-INK 664878, ERC-2018-AdG Mol-2D 788222), the Spanish MINECO (Project MAT2017-89993-R co-financed by FEDER and the Unit of Excellence "Maria de Maeztu" MDM-2015-0538), and the Generalitat Valenciana (Prometeo Programme) is gratefully acknowledged. S.G.M. and J.P.R. thank the Spanish MINECO for their predoctoral grants. M.C. acknowledges funding from the European Research Council (ERC) under the European Union's Horizon 2020 research and innovation programme (grant agreement No. 725767 - hyControl). Authors acknowledge Prof. Md. K. Nazeeruddin for providing the N965 molecules and Dr. H. Bolink for helpful discussions.

Received: ((will be filled in by the editorial staff))
Revised: ((will be filled in by the editorial staff))
Published online: ((will be filled in by the editorial staff))

## References

[1] D. Sun, E. Ehrenfreund, Z. Valy Vardeny, *Chem. Commun.* **2014**, *50*, 1781.

[2] D. M. Wolf, S. A., Awschalon, D.D, Buhrman, R.A., Daughton, R.A., von Molnár, S., Roukes, M.L., Chtchelkanova, A.Y., Treger, *Science* **2001**, *294*, 1488.

[3] I. Zutic, J. Fabian, S. Das Sarma, *Rev. Mod. Phys.* **2004**, *76*, 323.

[4] M. Cinchetti, V. A. Dediu, L. E. Hueso, *Nat. Mater.* **2017**, *16*, 507.

[5] J. Devkota, R. Geng, R. C. Subedi, T. D. Nguyen, *Adv. Funct. Mater.* **2016**, *26*, 3881.

[6] X. Sun, S. Vélez, A. Atxabal, A. Bedoya-Pinto, S. Parui, X. Zhu, R. Llopis, F. Casanova, L. E. Hueso, *Science* **2017**, *357*, 677.

[7] V. A. Dediu, L. E. Hueso, I. Bergenti, C. Taliani, *Nat. Mater.* **2009**, *8*, 707.

[8] J. Camarero, E. Coronado, *J. Mater. Chem.* **2009**, *19*, 1678.

[9] C. Boehme, J. M. Lupton, *Nat. Nanotechnol.* **2013**, *8*, 612.

[10] X. Sun, A. Bedoya-Pinto, Z. Mao, M. Gobbi, W. Yan, Y. Guo, A. Atxabal, R. Llopis, G. Yu, Y. Liu, A. Chuvilin, F. Casanova, L. E. Hueso, *Adv. Mater.* **2016**, 2609.

[11] E. Ehrenfreund, Z. Valy Vardeny, *Phys. Chem. Chem. Phys.* **2013**, *15*, 7967.

[12] T. D. Nguyen, E. Ehrenfreund, Z. V. Vardeny, *Science* **2012**, *337*, 204.

[13]  J. P. Prieto-Ruiz, S. G.Miralles, G. N., M. Aeschlimann, M. Cinchetti, H. Prima-García, E. Coronado, *Adv. Electron. Mater.* **2017**, *3*, 1600366.

[14]  Z. G. Yu, *Nat. Commun.* **2014**, *5*, 4842.

[15]  Z. H. Xiong, D. Wu, Z. V. Vardeny, J. Shi, *Nature* **2004**, *427*, 821.

[16]  E. Bergenti, I., Dediu, V. A., Arisi, C. Mertelj, T., Murgia, M., Riminucci, A., Ruani, G., Solzi, M., Taliani, *Org. Electron.* **2004**, *5*, 309.

[17]  G. Salis, S. F. Alvarado, M. Tschudy, T. Brunschwiler, R. Allenspach, *Phys. Rev. B* **2004**, *70*, 1.

[18]  M. Wohlgenannt, *Phys. Status Solidi - Rapid Res. Lett.* **2012**, *6*, 229.

[19]  M. Sessolo, H. J. Bolink, *Adv. Mater.* **2011**, *23*, 1829.

[20]  J. Chua, L. L., Zaumseil, E. C. W. Chang, J. F., Ou, H. Ho, P. K.H., Sirringhaus, R. H. Friend, *Nature* **2005**, *434*, 194.

[21]  P. Ruden, *Nat. Mater.* **2011**, *10*, 8.

[22]  A. Zhou, Yi., Fuentes-Hernandez, C., Shim, J., Meyer, J., Giordano, A. J., Li, H., Winget, P., Papadopoulos, T., Cheun, H., Kim, J., Fenoll, M., Dindar, A., Haske, W., Najafabadi, E., Khan, T. M., Sojoudi, H., Barlow, S., Graham, .S., Brédas, J. L., Marder,, B. Kippelen, *Science* **2012**, *873*, 327.

[23]  M. Piquemal-Banci, R. Galceran, S. Caneva, M. B. Martin, R. S. Weatherup, P. R. Kidambi, K. Bouzehouane, S. Xavier, A. Anane, F. Petroff, A. Fert, J. Robertson, S. Hofmann, B. Dlubak, P. Seneor, *Appl. Phys. Lett.* **2016**, *108*, 1.

[24]  S. S. P. Parkin, C. Kaiser, A. Panchula, P. M. Rice, B. Hughes, M. Samant, S.-H. Yang, *Nat. Mater.* **2004**, *3*, 862.

[25]  Desai, P., Shakya, P., Kreouzis, T., Gillin, W., Morley, N., *Phys. Rev. B* **2007**, *75*, 1.

[26] M. E. Macià, F., Wang, F., Harmon, N. J., Kent, A. D., Wohlgenannt, M., Flatté, *Nat. Commun.* **2014**, *5*, 3609.

[27] Ö. Mermer, G. Veeraraghavan, T. L. Francis, Y. Sheng, D. T. Nguyen, M. Wohlgenannt, A. Köhler, M. K. Al-Suti, M. S. Khan, *Phys. Rev. B - Condens. Matter Mater. Phys.* **2005**, *72*, 1.

[28] J. Wang, A. Chepelianskii, F. Gao, N. C. Greenham, *Nat. Commun.* **2012**, *3*, 1191.

[29] M. Yunus, P. P. Ruden, D. L. Smith, *Appl. Phys. Lett.* **2008**, *93*, 2006.

[30] L. Barraud, C., Seneor, P., Mattana, R., Fusil, S., Bouzehouane, K., Deranlot, C., Graziosi, P., Hueso, A. Bergenti, I., Dediu, V.A., Petroff, F., Fert, *Nat. Phys.* **2010**, *6*, 615.

[31] S. Zhang, X., Mizukami, M. Kubota, T., Ma, Q., Oogane, Y. Naganuma, H., Ando, T. Miyazaki, *Nat. Commun.* **2013**, *4*, 1392.

[32] S. Szulczewski, G., Sanvito, M. Coey, *Nat. Mater.* **2009**, *8*, 693.

[33] B. K. Markus Wohlgenannt, Peter A. Bobbert, *MRS Bull. (Organic Spintron.* **2014**, *39*, 590.

[34] V. A. Graziosi, P., Prezioso, M., Gambardella, A., Kitts, C., Rakshit, R.K., Riminucci, A., Bergenti, I., Borgatti, F., Pernechele, C., Solzi, M., Pullini, D., Busquets-Mataix, D., Dediu, *Thin Solid Films* **2013**, *534*, 83.

[35] E. Bolink, H. J., Baranoff, M. K. Clemente-León, M., Coronado, E., Lopéz-Muñoz, A., Repetto, D., Sessolo, M., Nazeeruddin, *Langmuir* **2009**, *25*, 79.

[36] R. Lin, F. Wang, J. Rybicki, M. Wohlgenannt, K. A. Hutchinson, *Phys. Rev. B - Condens. Matter Mater. Phys.* **2010**, *81*, 1.

[37] Y. Z. Wang, M. Yang, D. C. Qi, S. Chen, W. Chen, A. T. S. Wee, X. Y. Gao, *J. Chem. Phys.* **2011**, *134*, 1.